\documentstyle[twoside,fleqn,espcrc2,epsf]{article}

\newcommand{\AmS}{{\protect\the\textfont2
  A\kern-.1667em\lower.5ex\hbox{M}\kern-.125emS}}

\title{The QCD phase transition at high temperature and low density
\thanks{Presented by S. Ejiri. Numerical work was performed using the UKQCD 
APEmille in Swansea supported by PPARC grant PPA/A/S/1999/00026.
}}

\author{ S. Ejiri\rlap,\address{Department of Physics, University of 
Wales Swansea, Singleton Park, Swansea, SA2 8PP, U.K.} 
C.R. Allton\rlap,$^{\rm a}$ S.J. Hands\rlap,$^{\rm a}$ 
O. Kaczmarek\rlap,\address{Fakult\"{a}t f\"{u}r Physik, Universit\"{a}t 
Bielefeld, D-33615 Bielefeld, Germany.} F. Karsch\rlap,$^{\rm b}$ 
E. Laermann\rlap,$^{\rm b}$ and Ch. Schmidt$^{\rm b}$}

\begin{document}

\begin{abstract}
We study the thermal properties of QCD in the presence of a small quark 
chemical potential $\mu$. 
Derivatives of the phase transition point with respect to $\mu$ are 
computed at $\mu=0$ for 2 and 3 flavors of p-4 improved staggered fermions 
on a $16^3\times4$ lattice. 
%The resulting Taylor expansion is well behaved for 
%the small values of $\mu/T_c\sim0.1$ relevant for RHIC phenomenology, and 
%predicts a critical curve $T_c(\mu)$ in reasonable agreement with estimates
%obtained using exact reweighting. In addition, 
Moreover 
we contrast the case of isoscalar and isovector chemical
potentials, quantify the effect of $\mu\not=0$ on the equation of state, and 
comment on the screening effect by dynamical quarks and 
the complex phase of the fermion determinant in QCD with $\mu\not=0$.
\vspace{1pc}
\end{abstract}

\maketitle

%\section{Introduction}

To understand recent heavy-ion collision experiments, 
theoretical study of the QCD phase transition at high temperature 
and low density is important. 
For instance, the interesting regime for RIHC is $\mu_q/T_c \sim 0.1$, 
where $\mu_q = \mu /a$ is a quark chemical potential. 
However, the Monte-Carlo method is not directly applicable for simulations 
at $\mu \neq 0$, 
which makes the study of finite-density QCD difficult; 
hence we usually use the reweighting method. Using the identity 
\begin{eqnarray}
\label{eq:rew}
& \hspace{-3mm} \langle {\cal O} \rangle_{(\beta, \mu)} & \hspace{-3mm} 
%&=& \frac{1}{\cal Z} \int 
%{\cal D}U {\cal O} ( \det M(\mu))^{\alpha N_{\rm f}} {\rm e}^{-S_g}
%\\  && \hspace{-23mm}
= \left\langle {\cal O} W \right\rangle_{(\beta_0,0)} /
\left\langle W \right\rangle_{(\beta_0,0)}, \hspace{10mm} \\
 &\hspace{-3mm} W =& \hspace{-4mm} {\rm e}^{N_{\rm f}
(\ln \det M(\mu) - \ln \det M(0))} {\rm e}^{-S_g(\beta)+S_g(\beta_0)}, 
\nonumber
\end{eqnarray} 
the expectation value $\langle {\cal O} \rangle$ at $\mu \neq 0$ 
is computed by a simulation at $\mu=0$. 
Here $M$ is the fermion matrix, $S_g$ the gauge action, 
and $N_{\rm f}$ the number of flavors. 
%and $\alpha$ is 1 or $1/4$ for the Wilson or staggered fermion.
Then, there exists a famous ``sign problem''. 
Because $\det M$ is complex at $\mu \neq 0$, if the complex phase 
fluctuates rapidly, both numerator and denominator in RHS of 
eqn.(\ref{eq:rew}) become vanishingly small.
For the case of small $\mu$, the complex phase can be written 
by the odd terms of the Taylor expansion of $\ln \det M$ \cite{swan02}. 
Denoting $\det M = |\det M| {\rm e}^{i \theta}$, 
\begin{eqnarray}
\label{eq:phase}
\theta = N_{\rm f} \sum_{n: {\rm odd}} {\rm Im} 
\left( \frac{\partial^n \ln \det M}{\partial \mu^n} \right) \mu^n.
\end{eqnarray}
The first term is
$N_{\rm f} {\rm Im} \, {\rm tr} [M^{-1} (\partial M/\partial \mu)] \mu$. 
From these equations, we find explicitly that the magnitude of 
$\theta$ is proportional to $\mu$, the volume and $N_{\rm f}$. Moreover 
the first term of $\theta$
can be computed by the noise method and has a tendency that 
the phase fluctuation becomes larger as the quark mass or $T$ decreases.
Roughly speaking, the sign problem happens when the fluctuation of 
$\theta$ becomes larger than $O(\pi/2)$. 
Thus the sign problem is not serious for the range of small $\mu$, 
but that region becomes narrower and narrower as the volume increases, 
which suggests that in the thermodynamic limit, 
since the region where the sign problem is manageable decreases 
to zero size, the only way of accessing the $\mu \neq 0$ region is 
via a Taylor expansion since this involves calculating quantities
(i.e. derivatives) only at $\mu=0$.

Last year, we proposed a general formulation to compute 
the derivatives of physical quantities \cite{lat01}. 
We perform a Taylor expansion for $\ln \det M$ and fermionic operators 
in eqn.(\ref{eq:rew}) and neglect higher order terms of $\mu$. 
Then the resulting expectation value contains an error of 
higher order in $\mu$ but which does not affect the calculation 
of derivatives of lower order than the neglected terms.
In this report, we summarize the results obtained by this method.
We perform simulations on a $16^3 \times 4$ lattice using a combination 
of the Symanzik improved gauge and 2 flavors of the p4-improved 
staggered fermion actions \cite{p4action}. 
Details are written in Ref.~\cite{swan02}. 
We also comment on the $N_{\rm f}$ dependence, in the final section, 
by performing an additional 3 flavor simulation.

\begin{figure}[t]
%\vspace*{-9mm}
\centerline{
\epsfxsize=6.5cm\epsfbox{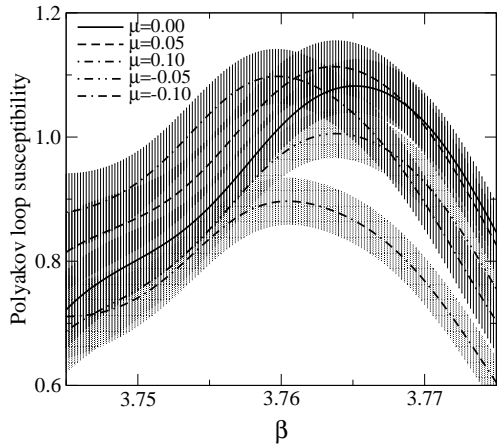}
}
\vspace*{-9mm}
\caption{
Polyakov loop susceptibility at $m=0.2$.
}
\vspace*{-2mm}
\label{fig:psu02}
\end{figure}

\paragraph{Phase transition line }
First of all, we discuss the phase transition line in the $(T, \mu)$ plane. 
Because the first derivative is expected to be zero from the symmetry 
under exchange of $\mu$ and $-\mu$, we calculate second derivative 
of $T_c$ with respect to $\mu$.
In Fig.~\ref{fig:psu02}, we plot Polyakov loop susceptibility 
as a function of $\beta$ for $\mu = 0, \pm0.05$ and $\pm0.1$ at $m=0.2$. 
This calculation contains errors at $O(\mu^3)$. 
From this figure, we find that the peak position of the susceptibility 
moves left as $\mu$ increases, which means that $\beta_c$ or $T_c$ 
decreases as $\mu$ increases.  
Assuming the peak position is at $\beta_c$, 
we determine the second derivative of $\beta_c$. 
Combining with results for the chiral susceptibility 
at $m=0.1$ and $0.2$, we obtain 
${\rm d}^2 \beta_c/{\rm d} \mu^2 \approx -1.1$ with $30$ - $50 \%$ error 
and any quark mass dependence of ${\rm d}^2 \beta_c/{\rm d} \mu^2$ 
is not visible within the accuracy of our calculation. 

\begin{figure}[t]
%\vspace*{-9mm}
\centerline{
\epsfxsize=6.5cm\epsfbox{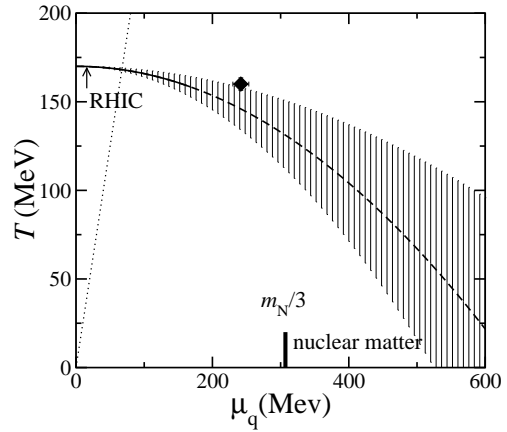}
}
\vspace*{-9mm}
\caption{
Sketch of the phase diagram. 
%, as estimated using our value of the curvature of $\beta_c(\mu=0)$. 
The diamond symbol is the critical point obtained by 
Fodor and Katz \protect\cite{Fod01}. 
Dotted line is upper bound of the fit range to determine the curvature.
}
\vspace*{-2mm}
\label{fig:cur}
\end{figure}

The second derivative of $T_c$ is given by 
\begin{eqnarray}
\frac{{\rm d}^2 T_c}{{\rm d} \mu_{\rm q}^2} = -\frac{1}{N_t^2 T_c} 
\left. \frac{{\rm d}^2 \beta_c}{{\rm d} \mu^2} \right/ 
\left( a \frac{{\rm d} \beta}{{\rm d} a} \right),
\end{eqnarray}
with 
$a({\rm d} \beta / {\rm d} a)$ obtained from the string tension data in 
\cite{p4action}. 
%We get 
%$a^{-1}({\rm d}a/{\rm d}\beta)=-2.08(43)$ at $(\beta,m)=(3.65,0.1)$. 
We then find 
$T_c({\rm d}^2 T_c/{\rm d}\mu^2_{\rm q}) \approx -0.14$. 
We sketch the phase transition line from the curvature with 
$50 \%$ error in Fig~\ref{fig:cur}.
%assuming $T_c \simeq 170{\rm MeV}$. 
At the relevant point for RHIC, this shift of $T_c$ is very small 
from that at $\mu=0$ 
and the result is roughly consistent with those obtained by 
other groups \cite{Fod01,Phil}.

\paragraph{Screening effect by dynamical quarks}
Under the presence of chemical potential, the time reversal symmetry 
is broken. 
By the asymmetry, an interesting property is shown 
in the measurement of the Polyakov loop, 
%that is interpreted as 
an external quark current running in the positive time direction.  
Negative chemical potential induces the dynamical generation of 
anti-quarks, which in contrast to quarks can completely screen 
an external color triplet current. 
Thus the free energy of a single quark is reduced, especially 
in the confinement phase, 
and the singularity at the phase transition point is weakened, 
since the long range fluctuation is screened.
This effect can be seen in Fig.~\ref{fig:psu02},  where 
we denote the Polyakov loop susceptibility $\chi_L$ at $\mu<0$ by 
dot-dot-dash and dot-dash-dash lines. 
We see that the peak height of $\chi_L$ becomes smaller for $\mu<0$ 
corresponding to a weaker singularity, while
the peak position is almost the same between positive and negative $\mu$. 
%The screening effect only seems to make the phase transition singularity 
%weaker without shifting the pseudocritical line. 
Simultaneously, we observe that the Polyakov loop at $\mu < 0$ is 
larger than that at $\mu >0$, suggesting the free energy of 
a single quark is reduced.
Note that this property cannot be seen in 2-color QCD 
where systems at $\mu$ and $-\mu$ are identical.

\paragraph{Equation of state}
Next we discuss the $\mu$-dependence of the equation of state 
which describes the energy density $\epsilon$ and pressure $p$. 
If we employ the integral method based on the homogeneity of the system, 
we obtain $p=-f$, where $f=-(T/V)\ln{\cal Z}$.
Then derivatives of $p$ with respect to $\mu$ are related to
the quark number density $n_{\rm q}$ 
and the singlet quark number susceptibility
$\chi_{\rm S}=\partial n_{\rm q}/\partial\mu_{\rm q}$
\cite{Gott88}: 
\begin{eqnarray}
\label{eq:eosder1}
\frac{\partial (p/T^4)}{\partial \mu_{\rm q}} 
%= \frac{\partial (-f/T^4)}{\partial \mu_{\rm q}} 
%= \frac{1}{VT^3} \frac{\partial \ln {\cal Z}}{\partial \mu}_{\rm q} 
= \frac{n_{\rm q}}{T^4}, \ \ 
% \\ \label{eq:eosder2}
\frac{\partial^2 (p/T^4)}{\partial \mu_{\rm q}^2} 
%&=& \frac{\partial^2 (-f/T^4)}{\partial \mu_{\rm q}^2} 
%= \frac{1}{VT^3} \frac{\partial^2 \ln {\cal Z}}{\partial \mu_{\rm q}^2} 
= \frac{\chi_{\rm S}}{T^4}. 
\end{eqnarray}
The quark number density is zero at $\mu=0$ so the leading correction 
is $O(\mu^2)$.
Moreover the second derivative of $\epsilon$ can also be estimated by
\begin{eqnarray}
\label{eq:e-3pd2}
\frac{\partial^2 (\epsilon -3p)/T^4}{\partial \mu_{\rm q}^2} 
\approx - \frac{1}{T^4} \frac{\partial \chi_{\rm S}}{\partial \beta}
 \left( \frac{1}{a} \frac{\partial a}{\partial \beta} \right)^{-1}.
\end{eqnarray} 
Here we neglect $a(\partial m/\partial a)$, an approximation 
which is valid in the chiral limit.
We obtain
$T^2 \partial^2(p/T^4)/\partial \mu_{\rm q}^2 \approx 0.69$ and 
$T^2 \partial^2 (\epsilon /T^4)/\partial \mu_{\rm q}^2 \approx 10.6$ 
at $\beta_c$ for $m=0.1$.
The discrepancy of $p/T^4$ ($\epsilon /T^4$) in the RHIC regime 
$\mu_{\rm q}/T_c \sim 0.1$ from its value at $\mu=0$ 
is about $0.0035$ ($0.05$). 
This is a 1\% effect, and hence quite small. 

From the second derivatives of $p$ and $\epsilon$ with respect to 
$\mu$ together with the derivatives with respect to $T$, 
we calculate the line of constant pressure and energy density.
%, $\Delta p(T,\mu^2)=0$ and $\Delta \epsilon(T,\mu^2)=0$. 
%The slope of the constant line is written by
%\begin{eqnarray}
%\frac{{\rm d} T}{{\rm d} (\mu^2)} = \left. 
%- \frac{\partial(p/T^4)}{\partial (\mu^2)} 
%\right/ \left( \frac{\partial (p/T^4)}{\partial T} + \frac{4p}{T^5} \right).
%\end{eqnarray}
We find that the slope of the constant pressure (energy density) line is 
$T({\rm d} T / {\rm d} (\mu_{\rm q}^2)) \approx -0.11 (-0.09).$ 
As the slope of $T_c$ in $\mu_q^2$ is 
$T_c ({\rm d} T_c / {\rm d} (\mu_{\rm q}^2)) 
%= (1/2) T_c ({\rm d}^2 T_c / {\rm d} \mu_{\rm q}^2) 
\approx -0.07$,
this result suggests that the line of constant pressure or 
energy density is parallel with the phase transition line.

\paragraph{Iso-vector chemical potential}
If instead we were to 
impose an {\sl isovector\/} chemical potential $\mu_I$ 
having opposite sign for $u$ and 
$d$ quarks \cite{SS}, 
then the quark determinant would become real and positive, enabling
simulations using standard Monte-Carlo methods \cite{KogSin}.
%This motivates a comparison between systems with the usual isoscalar
%chemical potential $\mu$ and the isovector chemical potential $\mu_I$.
In the framework of the Taylor expansion, terms even in $\mu$ are identical
for both $u$ and $d$ quarks, but odd terms 
cancel for the case $\mu_I\not=0$.
We analyzed the transition point $\beta_c(\mu_I)$ and 
do not observe significant difference for $\beta_c$ between 
$\mu$ and $\mu_I$ in the region of small $\mu$, 
which is different from a naive expectation that the phase transition 
line for $\mu_I$ runs toward the $T=0$ onset threshold of a pion condensate 
at a critical $\mu_{Io}\simeq m_{PS}/2<m_N/3$, 
and hence the curvature for $\mu_I$ larger than for isoscalar $\mu$. 
However the quark mass we used is still large and $m_{\pi}$ is 
not so small. Simulations at small $m$
are necessary to check the naive picture.

\paragraph{$N_{\rm f}$ dependence}
Finally, we comment on the difference between $N_{\rm f}=2$ and $3$. 
We performed an additional simulation of $N_{\rm f}=3$.
% on a $16^3 \times 4$ lattice.
The preliminary results of the curvature at $m=0.1$ are 
$T_c({\rm d}^2 T_c/{\rm d}\mu^2_{\rm q}) \approx -0.15$ and $-0.11$ 
for the cases $\mu_{ud}=\mu_s=\mu_{\rm q}$ and 
$\mu_{ud}=\mu_{\rm q}, \mu_s=0$ respectively.
There is no significant difference from $N_{\rm f}=2$.  
Therefore we expect that our result for $N_{\rm f}=2$ is not 
so different from real QCD lying
between $N_{\rm f}=2$ and $3$.

The most interesting point for $N_{\rm f}=3$ is the existence 
of a critical quark mass $m_c$ on the $\mu=0$ axis 
between a first order phase transition at small $m$ and 
a crossover at large $m$. 
We also expect such a critical point at $\mu \neq 0$ even 
in the region of large $m$. Hence to investigate the relation 
between these critical points is quite important. 
The first preliminary result is reported by \cite{Schm02}.

\end{document}